# Ultrafast Oscillations of a Ballistically Propagating Polariton Condensate Driven by Inter-mode Coherent Energy Transfer

Fangxin Zhang,[1,*] Changchang Huang,[1,*] Yichun Pan,[1] Guangran Yang,[1] Wei Xie,[2] and Weihang Zhou[1,†]

[1] *Wuhan National High Magnetic Field Center and School of Physics,*
*Huazhong University of Science and Technology, Wuhan, Hubei 430074, China*

[2]*State Key Laboratory of Precision Spectroscopy, School of Physics and Electronic Science,*
*East China Normal University, Shanghai 200241, China*

(Dated: December 4, 2025)

The interplay of dynamics and transport leads to intriguing spatiotemporal behaviors of nonequilibrium macroscopic quantum systems. By means of time-resolved spectroscopy, we here provide microscopic insights into the interplay of ballistic transport and many-particle interactions for an exciton-polariton condensate. We observed anomalous condensation of a coherent polariton flow propagating away from the hot reservoir, accompanied by ultrafast oscillations of its population in the time domain with a period of a few picoseconds. On the basis of time- and spatially-resolved photo luminescence imaging characterization, we reveal that the inter-mode coherent energy transfer, controllable via an incoherent excitonic reservoir, gives rise to the observed ultrafast oscillations. Theoretically, modeling was conducted using an open-dissipative Gross-Pitaevskii equation, with the simulation results fully reproducing our experimental observations. These results advance the fundamental understanding of light-matter interactions in nonequilibrium systems associated with the interplay of dynamics and transport.

Exciton-polaritons, as bosonic quasi-particles with hybrid light-matter nature, have been intensively studied as an ideal platform for the study of macroscopic quantum phenomena [1-6]. They show striking similarities to Bose-Einstein condensates of atomic gases, such as super-fluidity [7-9] and vortices [10-12]. However, due to their nonequilibrium condensation nature, excitonpolaritons have demonstrated peculiarities of their own that set them apart from conventional atomic condensates. Thanks to the efforts devoted in the past decades, it is nowadays clear that such peculiarities arise from their distinctive half-light half-matter nature. On one hand, their inherent photonic component endows them with ultrashort lifetimes of the order of a few picoseconds, naturally placing them in nonequilibrium dissipative systems that exhibit rich dynamic behavior. Indeed, a variety of dynamical behaviors, such as relaxation oscillations [13-15], parametric scattering [16-19], and excitedstate condensation [20, 21], have been observed in various microcavity and material systems, showcasing the rich physics of polaritons. On the other hand, the inherent excitonic component introduces significant inter-particle interactions, giving rise to fascinating polariton nonlinearities. Notably, intricate transport behavior can thus be triggered [22-25]. From a device-application perspective, hybridization of excitons with cavity photons may overcome the inefficient and diffusive exciton transport in semiconductors, which could find applications in a variety of fields, such as organic photovoltaics [26, 27] and excitonic devices [28, 29]. However, while both polariton dynamics and their transport receive much attention in the past decades, there is a lack of fundamental understanding of the intricate spatiotemporal behaviors associated with the interplay of polariton dynamics and transport. In this work, we aim to provide microscopic insights into the interplay of ballistic transport and many-particle interactions for an exciton-polariton condensate. We tracked the relaxation and transport of a ballistically propagating polariton condensate, formed by nonresonant optical excitation, by means of time-resolved spectroscopy and angle-resolved photo-luminescence imaging technique. Through power-dependent studies, we observed an anomalous accumulation of population at the ground state for a coherent polariton flow propagating at high velocities. Interestingly, ultrafast oscillation of the polariton population was observed by our streak camera measurements, accompanying the anomalous condensation behavior. By combining time-and angle-resolved analyses, we successfully reveal that the inter-mode coherent energy transfer, controllable via an incoherent excitonic reservoir, is responsible for the observed ultrafast oscillation. The proposed model was further confirmed by our opendissipative Gross-Pitaevskii equation simulation which fully reproduces the observed oscillations.

The samples we used are one-dimensional (1D) ZnO microrods synthesized by means of chemical vapor deposition method. They typically have diameters of 1-3 $\mu m$ and lengths of up to 100 $\mu m$. Due to their regular hexagonal cross-sections, such micro-rods form naturally whispering gallery micro-cavities and are very good platforms for the study of strong excitonphoton coupling, as demonstrated in our previous work [14, 30, 31]. The light source we used is a 370 $nm$ femtosecond laser with pulse width of 300 $fs$ and repetition rate of 200 kHz. Population distribution of polaritons in momentum space is detected by a home-built angle-resolved photoluminescence imaging system. Time-resolved spectroscopy is carried out using a streak camera (Hamamatsu, C10910) with a time resolution of ~ 2ps. An adjustable confocal hole is introduced into the optical system to provide spatial resolution for angle-resolved imaging and time-resolved measurements. All optical measurements were carried out at room temperature.

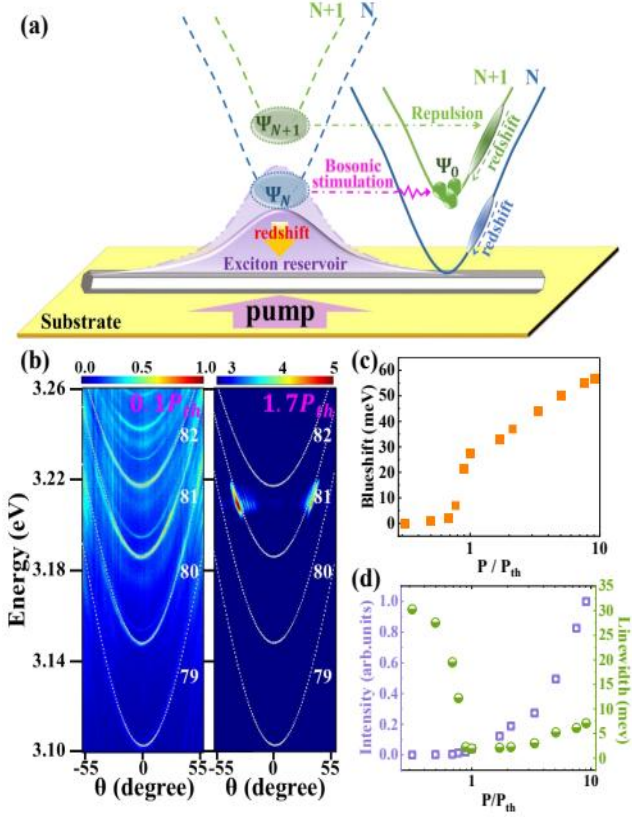

FIG. 1. Physics model and photoluminescence characterization. (a) Schematic diagram for the ultrafast oscillations of a ballistically propagating polariton condensate driven by intermode coherent energy transfer. (b) Typical angle-resolved photoluminescence images of the as-synthesized ZnO microrod. Left image: pumping power ~ 0.1 $P_{th}$; Right image: pumping power ~ 1.7 $P_{th}$. Numbers in the image denote the mode order of the polariton dispersions. (c) Blueshift of the 81st branch as a function of pumping power. (d) Emission intensity (purple open squares) and linewidth (green half-filled circles) of the 81st branch as a function of pumping power.

Typical angle-resolved photoluminescence images of the as-synthesized ZnO microrods are shown in Fig. 1(b). The image in the left shows the result at low pumping power (~ 0.1 $P_{th}$, with Pth being the threshold power for polariton lasing). A series of well-defined parabolic emission belts, which correspond to the lower polariton branches, are clearly visible. The dotted curves together shown in the image are theoretical fittings to the observed emission belts using the coupled-oscillator model. The strong coupling regime can be confirmed by the excellent agreement between the theoretical curves and the experimental results. The right image in Fig. 1(b) shows the typical above-threshold angle-resolved measurement result. In contrast to conventional polariton condensation where polaritons are expected to accumulate at the bottom of the dispersion curve, two bright spots are observed to emerge along the dispersion curve of the 81st branch in the large angle region (~ 34.5°). The physical picture behind this phenomenon is as follows: driven by the repulsive force of the heavy exciton reservoir, the optically injected polariton condensate propagates away from the center of the pumping spot. During this process, the condensate gains momentum (and kinetic energy) at the expense of its potential energy. At the edge of the excitation spot, momentum of the polariton condensate reaches its maximum value, as the excess potential energy from polariton-reservoir interaction vanishes at the edge of the pumping spot. The vanishing polariton emission around 0° in the angle-resolved image is a direct manifestation of the strong repulsion of polaritons by the exciton reservoir. Fig. 1(c) shows the power dependence of the energy shift of the 81st polariton dispersion, which is also a manifestation of the pronounced polariton-reservoir interactions. Fig. 1(d) shows the power dependence of the emission intensity and linewidth. Threshold behaviors for both curves can be clearly observed, which are typical for polariton condensations.

To further reveal the nonlinear effects of polaritons in the high-density regime, we continued our measurements at higher pumping power. Typical angle-resolved photoluminescence images under high pumping power are plotted in Fig. 2. In addition to the bright spots locating at ± 34.5°, new features emerge when pumping powers are increased to ~3.0$P_{th}$. As one can see from Fig. 2(c), for pumping power $P \approx 3.4 P_{th}$ a new spot appears at the bottom of the 81st dispersion curve, suggesting creation of polariton condensate with nearly zero in-plane momentum. Interestingly, similar behaviors can also be found for the 80th branch. As one can see from Fig. 2(c), large-angle bright spots of the 80th branch start to emerge for pumping powers of ~3.4 $P_{th}$. However, by further increasing the pumping power to~7.5$P_{th}$, condensation signals at the dispersion minimum of the 80th branch are clearly visible, as demonstrated in Fig. 2(e).

The anomalous condensation of propagating polaritons at the minimum of the dispersion curves cannot be rationalized using the potential-to-kinetic energy conversion picture as employed in previous reports [32, 33]. To unravel the underlying mechanism, we carried out timeresolved photoluminescence spectroscopic studies using a streak camera. Typical power dependence of the timeresolved photoluminescence images is shown in Fig.3. As one can see, what we observe under belowthreshold pumping is the spontaneous decay of the emission intensity, as expected. When pumping power was increased to the condensation threshold, polaritons show an accelerated decay due to the Bosonic stimulation, as one can see from Figs. 3(d)–3(e). From the power-dependent images, it can also be seen clearly that polaritons condense sequentially from high energy branch to low energy branch. More interestingly, under pumping power of $P \approx 3.4 P_{th}$, oscillations can be observed for the 81$^{st}$ branch, as one can see from Fig. 3(g). These oscillations become more and more pronounced as pumping power was further increased, as shown in Figs. 3(h)–3(j). Recalling the fact that the anomalous condensation at the minimum of the dispersion curve also emerges at pumping power $P \approx 3.4 P_{th}$ (Fig. 2(c)), we speculate that the observed oscillations manifest the anomalous condensation in the time domain. This speculation can be confirmed by the power-dependent behaviors of the 80$^{th}$ branch. As one can see from Fig. 2(e), the anomalous condensation for the 80$^{th}$ branch emerges at pumping power of ~7.5$P_{th}$. On the other hand, we found that oscillation behaviors for the 80th branch also emerge at pumping power of ~7.5$P_{th}$, as one can see from Fig. 3(i).

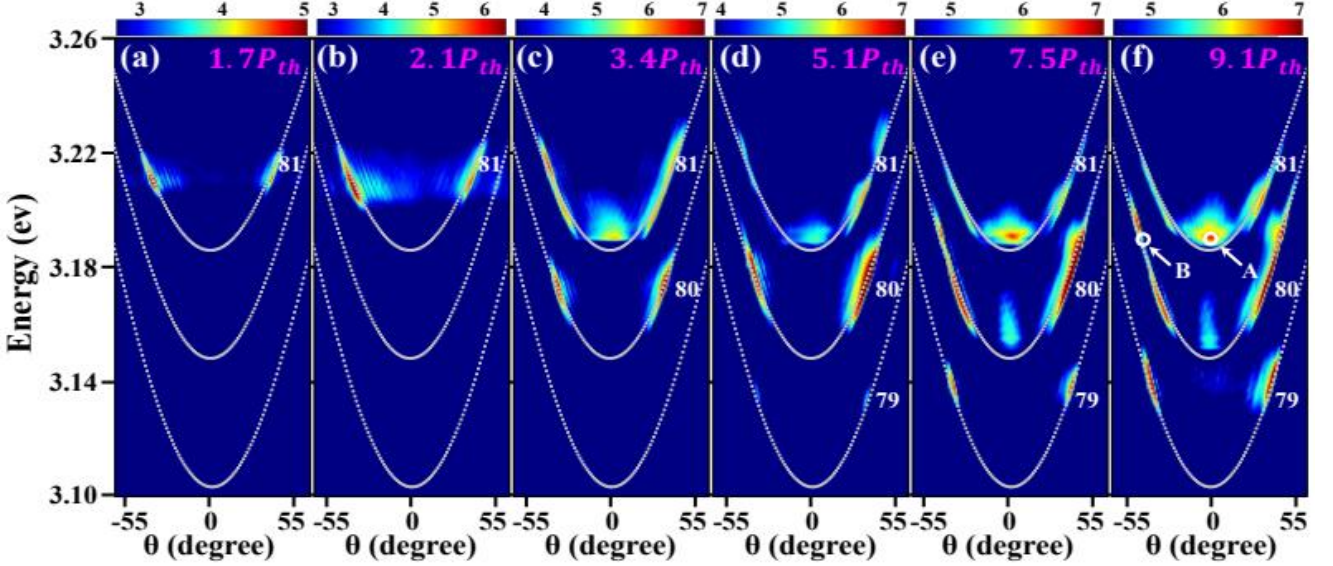

FIG. 2. Angle-resolved photoluminescence images of 1D ZnO microrod under high pumping power. (a) Pumping power ~ 1.7$P_{th}$, (b) 2.1$P_{th}$, (c) 3.4$P_{th}$, (d) 5.1$P_{th}$, (e) 7.5$P_{th}$, and (f) 9.1$P_{th}$. The letters A and B in (f) denote positions selected for further signal analyses, as discussed in the main text.

Based on our observations and the intrinsic characteristics of polaritons, we propose that the anomalous condensation in momentum space and the accompanying oscillations in time domain stem from the intermode coherent energy transfer. The physical picture of our model is shown schematically in Fig. 1(a). Due to the pronounced repulsive interaction, the excitonic reservoir essentially acts as a potential barrier for excitonpolaritons. The optically injected polariton condensate, which initially locates at the spot center (the top of the potential hill), propagates away due to Coulomb repulsion and gains momentum during its propagation at the expense of its potential energy. Its momentum reaches the maximum value at the edge of the pumping spot. However, due to the decaying population of excitons in the reservoir, energy of the polariton condensate in the spot center experiences continuous redshift. In cases that the initial blueshift of the condensate is larger than the energy gap between neighboring polariton branches ($\Delta E = E_{N+1} - E_N$), there exists a special moment when condensate of the Nth branch in the spot center is resonant in energy with the dispersion minimum of the $(N+1)^{th}$ branch at the spot edge during their relaxation process, as shown schematically in Fig. 1(a). In this case, stimulated scattering of polaritons from the Nth branch to $(N+1)^{th}$ branch could be triggered, as a result of their bosonic nature of polaritons. Due to the high efficiency of bosonic stimulation, there would be a sudden increase in the originally decaying polariton population of the $(N+1)^{th}$ branch. If we trace the population of the $(N+1)^{th}$ branch in the time domain, we would be able to observe oscillatory behaviors in its decaying curve.

To verify our model, we carried out theoretical simulations using the opendissipative Gross–Pitaevskii (GP) equations. To reduce the complexity of the simulations while preserving the essential characteristics of the system, we assume that excitonpolaritons in the pumping spot are repelled to the edge of the spot immediately upon condensation. Under this assumption, our proposed model can be described by the following GP equations:

$$i\hbar\frac{\partial \Psi_{N+1}}{\partial t} = \left[-\frac{\hbar^2}{2m}\nabla^2 + g|\Psi_{N+1}|^2 + g_R N_R \right.$$
$$\left. +\frac{i\hbar}{2}(\beta_{N+1}N_R - \gamma_{N+1})\right]\Psi_{N+1}, \quad (1)$$

$$i\hbar\frac{\partial \Psi_N}{\partial t} = \left[-\frac{\hbar^2}{2m}\nabla^2 + g(|\Psi_N|^2 + |\Psi_0|^2) + g_R N_R \right.$$
$$\left. +\frac{i\hbar}{2}(\beta_N N_R - \gamma_N - \Gamma|\Psi_0|^2)\right]\Psi_N, \quad (2)$$

$$i\hbar\frac{\partial \Psi_0}{\partial t} = \left[-\frac{\hbar^2}{2m}\nabla^2 + g(|\Psi_N|^2 + |\Psi_0|^2) \right.$$
$$\left. +\frac{i\hbar}{2}(\Gamma|\Psi_N|^2 - \gamma_0 -)\right]\Psi_0, \quad (3)$$

$$\frac{\partial N_R}{\partial t} = -\gamma_R N_R - (\beta_{N+1}|\Psi_{N+1}|^2 + \beta_N|\Psi_N|^2)N_R$$
$$+P(x)e^{-\gamma_P t}. \quad (4)$$

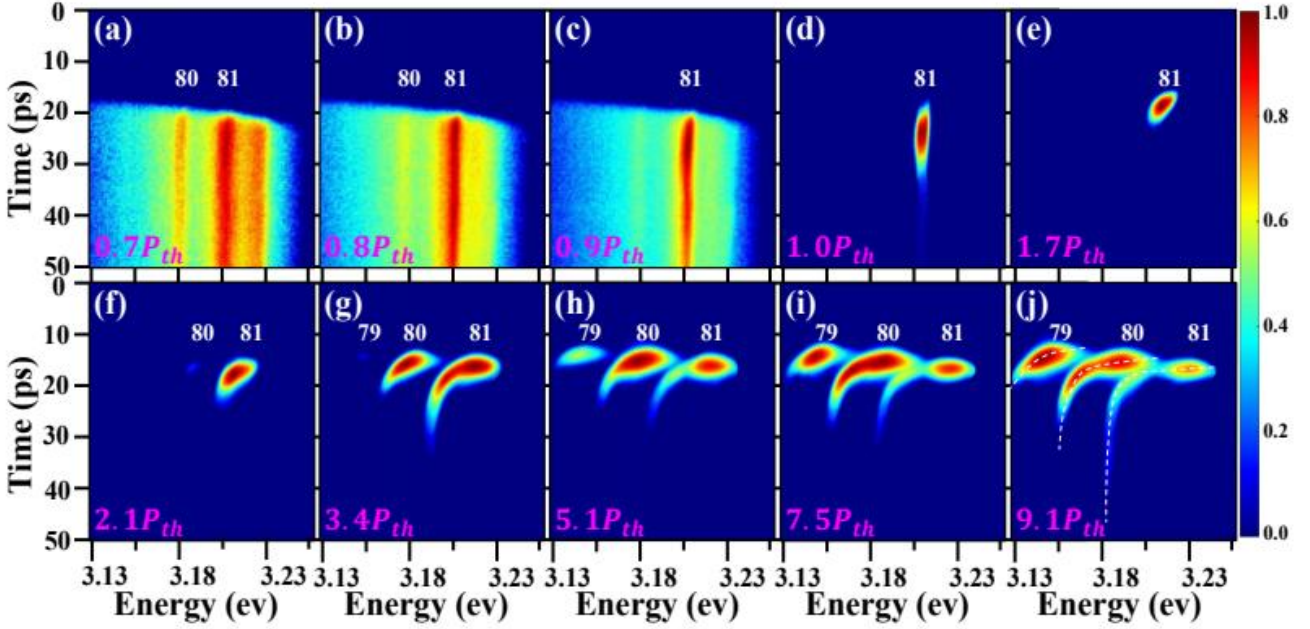

FIG. 3. Time-resolved photoluminescence images of the ZnO microrod. (a) Pumping power ~ $0.7P_{th}$; (b) $0.8P_{th}$; (c) $0.9\,P_{th}$; (d) $1.0P_{th}$; (e) $1.7P_{th}$; (f) $2.1P_{th}$; (g) $3.4P_{th}$; (h) $5.1P_{th}$; (i) $7.5P_{th}$; (j) $9.1P_{th}$. The white numbers in the image denote the mode order of corresponding polariton branches. The white dashed curves in (j) highlight the evolution traces of polaritons from corresponding branches in the time domain.

Here, evolution of polariton condensates for the $(N+1)^{th}$ and $N^{th}$ branches is described by Eqs. (1)–(2), respectively. Eq. 3 describes the evolution of polariton condensate at the dispersion minimum of the $(N+1)^{th}$ branch, which is replenished via stimulated scattering of polaritons from the $N^{th}$ branch, as depicted in Fig. 1(a). Eq. 4 describes the evolution of the excitonic reservoir. $\Psi_{N+1}$ and $\Psi_N$ represent wavefunctions of polaritons for the $(N+1)^{th}$ and $N^{th}$ branches, respectively. $\Psi_0$ corresponds to wavefunction of polaritons scattered from the $N$th branch to the dispersion minimum of the $(N+1)^{th}$ branch. $N_R$ is the number of excitons in the reservoir. $g$ and $g_R$ refer to the coefficients of polariton-polariton and polariton-reservoir interactions, respectively. $\beta_N$ ($\beta_N+1$) is the stimulated scattering rate between the reservoir and the $N^{th}((N+1)^{th})$ branch. $\Gamma$ refers to the stimulated scattering rate of polariton condensate from the $N^{th}$ branch to the dispersion minimum of the $(N+1)^{th}$ branch. $\gamma_N$, $\gamma_{N+1}$ and $\gamma_0$ represent the decay rate for corresponding polariton states. $P(x)$ describes the Gaussian spatial profile of the pumping laser. And $\gamma_P$ is the decay rate of the laser pulse.

The simulated temporal evolution curves of the emission intensities for the 79th, 80th, and 81st branches under pumping power of $P \approx 3.4P^{th}$ are plotted by the solid curves in Fig. 4(c). As one can see, our simulations fully reproduce the experimentally observed oscillations for the 81st branch. Meanwhile, satisfactory agreements for the 79th and 80th branches can also be found. Fig. 4(d) shows the comparison between theoretical calculations and the experimental results under higher pumping power of ~ $9.1P_{th}$, where oscillations can be observed for both 80th and 81st branches. As demonstrated, excellent agreements are again found, giving strong support to our proposed model.

Till now, we have confirmed through theoretical calculations that the observed oscillations stem from the inter-mode stimulated scattering process. Here, it is noteworthy that energetic resonance is a prerequisite to trigger such inter-mode stimulated scattering. To further confirm our model, we extracted the temporal evolution of the blueshift of polariton condensate at the 81st branch from its streak camera image. Typical result under pumping power of ~ 9.1 $P_{th}$ is plotted in Fig. 4(a). As one can see, blueshift of the polariton condensate at the 81st branch falls to be the same as the energy gap between the 80th and 81st branches at t ≈ 14.5 $ps$. The data in Fig. 4(d) shows that the second emission peak of the 81st branch, arising from the inter-mode stimulated scattering, appears at ~ 18.2 $ps$. This coincides with the time at which the energy resonance condition is met, within the margin of error. On the other hand, as a nonlinear effect, the proposed inter-mode stimulated scattering is expected to be strongly power density dependent. To check this issue, we tried to extract the power dependence of the scattering efficiency. As shown in Fig. 2(f), two representative points (A & B) in the angle-resolved photoluminescence image were selected for study. Point A corresponds to polariton condensate at the dispersion minimum of the 81st branch. Point B corresponds to polariton condensate from the 80th branch but with the same energy as those from Point A. As an approximation, the intensity ratio of A to B could be used as a probe to reveal the power-dependent characteristics of the intermode stimulated scattering. Typical power dependence of $I_A / I_B$ is shown in Fig. 4(b). As demonstrated, threshold behavior can be clearly identified, which is again in satisfactory agreement with our expectation.

Besides the evidences discussed above, fingerprints supporting our model can also be found from spatially- and time-resolved spectroscopic imaging analyses. As depicted in Fig. 1(a), the proposed inter-mode stimulated scattering is expected to occur primarily at the edge of the pumping spot for the following two reasons: (1) Polaritons at the edge have the largest energy difference from those at the center, making the energy resonance condition easier to satisfy. (2) At the edge of the spot, polaritons scattered to the dispersion minimum can maintain a zero group velocity state. This facilitates their collisions with other propagating polaritons, resulting in a more efficient stimulated scattering effect. Following these ideas, we carried out spatially- and time-resolved photoluminescence imaging measurements. Typical results, measured from another detection position of the same ZnO microrod as those shown in Figs. 1-3, are plotted in Figs. 4(e)-4(g). For comparison, Fig. 4(e) shows a time-resolved photoluminescence image without spatial filtering of the emission signals. Oscillation of the emission can be clearly identified, as highlighted by the white dashed curve. Fig. 4(f) shows the time-resolved image taken from the central region of the pumping spot only. Strikingly, oscillation of the emissions disappears completely. In Fig. 4(g), we show the time-resolved image taken at the edge of the pumping spot. As one can see, clear oscillations of the emission signals are again visible, in full agreement with our expectation. These excellent consistencies confirm again the validity of our proposed model.

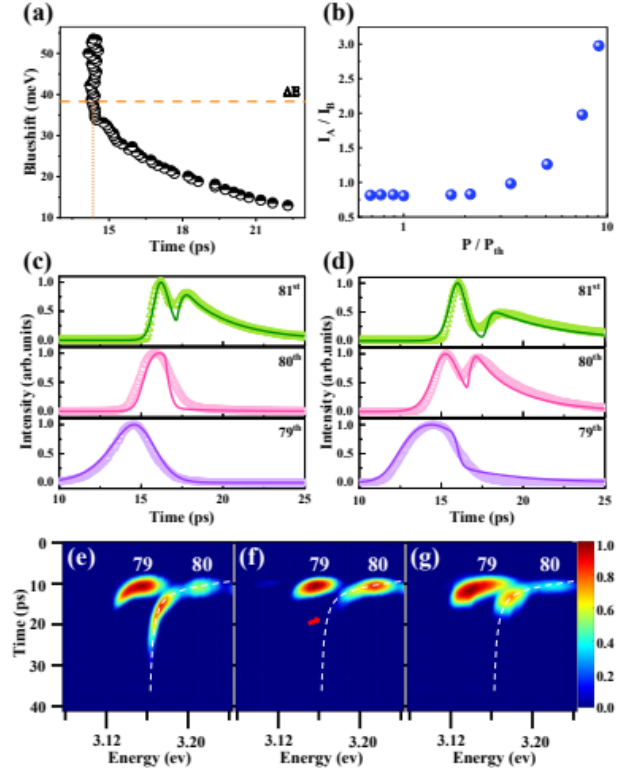

FIG. 4. Confirmation of the mechanism. (a) Temporal evolution of the blueshift of polariton condensate at the 81st branch under pumping power of ~ 9.1 $P_{th}$. $\Delta E$ refers to the energy gap between the 80th and 81st branches. (b) Ratio of the emission intensities from the two selected regions A and B in Fig. 2(f) as a function of the pumping power. (c) Temporal evolution of the emission intensity for the 79th, 80th, and 81st branches under pumping power of ~ 3.4 $P_{th}$. Solid curves are simulated results from our GP equation model. (d) Temporal evolution of the emission intensities under pumping power of ~ 9.1 $P_{th}$ and the corresponding simulation results (solid curves). The intensity traces are extracted along the white dashed curves shown in Fig. 3(j). (e) Time-resolved photoluminescence image from another detection position of the same ZnO microrod as those shown in Figs. 1-3. No spatial filtering of the emission signals was applied. (f) Time-resolved photoluminescence image taken from the central region of the pumping spot only. The red arrow is introduced to highlight the disappearance of the oscillations. (g) Time-resolved photoluminescence image taken at the edge of the pumping spot. The white dashed curves in (e-g) are introduced to highlight the evolution trace of polaritons from the 80th branch.

In summary, we report experimental observation of ultrafast oscillations of a ballistically propagating polariton condensate driven by inter-mode coherent energy transfer. We unraveled the intriguing characteristics of these ultrafast oscillations through angle-resolved spectroscopy and time-resolved photoluminescence imaging technique. The underlying mechanism was confirmed by combining theoretical simulations and spatially-resolved ultrafast spectroscopy. In contrast to conventional oscillations in a condensate, our observations highlight the importance of the interplay between dynamics and transport in polariton physics. These findings could thus advance the fundamental understanding of light-matter interactions in nonequilibrium systems and stimulate development of polariton-based optoelectronic devices.

This work was financially supported by the National Natural Science Foundation of China (Grant No. 12274159), and the National Key Research and Development Program of China (Grant No. 2022YFA1602700).

* These authors contributed equally to this work.
† zhouweihang@hust.edu.cn